\documentstyle[preprint,aps]{revtex}
\begin{document}
\title{Reply to the comment on `` Manipulating the frequency-entangled states by an
acoustic-optical modulator''}
\author{B.-S Shi, Y.-K Jiang and G.-C Guo}
\address{Lab. of Quantum Communication and Quantum Computation, Department of Physics%
\\
University of Science and Technology of China\\
Hefei, 230026, P. R. China}
\maketitle

\begin{abstract}
In the paper [1], authors claim that the scheme for entanglement swapping
using an acoustic-optical modulator [2] is flaw. In this reply, we show
there is a trivial mistake in the scheme [2], but the main conclusion is
still correct, that is, the acoustic-optical modulator can be used to
manipulate the frequency-entangled state.
\end{abstract}

In the paper [1], Resch et.al show that the transformation done by
acoustic-optical modulator (AOM) should be

\[
\left| \omega \right\rangle _1\stackrel{AOM}{\rightarrow }\frac 1{\sqrt{2}}%
[\left| \omega \right\rangle _t-i\left| \omega +\delta \right\rangle _d] 
\]

\begin{equation}
\left| \omega +\delta \right\rangle _{1^{^{\prime }}}\stackrel{AOM}{%
\rightarrow }\frac 1{\sqrt{2}}[\left| \omega \right\rangle _t+i\left| \omega
+\delta \right\rangle _d],
\end{equation}
(where $t,d,1,1^{^{\prime }}$ refer to the special mode labels shown in Fig
1 of the Ref. [1], $\omega ,\omega +\delta $ are photon frequencies), not be

\[
\left| \omega \right\rangle _1\stackrel{AOM}{\rightarrow }\frac 1{\sqrt{2}}%
[\left| \omega \right\rangle _t+\left| \omega +\delta \right\rangle _d]
\]
\begin{equation}
\left| \omega +\delta \right\rangle _{1^{^{\prime }}}\stackrel{AOM}{%
\rightarrow }\frac 1{\sqrt{2}}[\left| \omega \right\rangle _t+\left| \omega
+\delta \right\rangle _d].
\end{equation}
This is true. We make a trivial mistake in this transformation, and we thank
them very much for pointing out this mistake. But this small mistake does
not change our conclusion, that is, AOM can be used to manipulate the
frequency-entangled state. Next, we show it in detail.

In the Ref. [1], authors claim that no entanglement swapping has occurred
between photons 1 and 4, according to Eq. (12) of the Ref.[1], which is
derived from the proper transformation of Eq. (1) done by AOM. But Eq.[12]
can be rewritten as the following

\begin{eqnarray}
&&\left| \omega +\delta \right\rangle _{1^{^{\prime }}}\left| \omega +\delta
\right\rangle _4(\left| \omega \right\rangle _{T_2}-i\left| \omega +\delta
\right\rangle _{T_2^{^{\prime }}})(\left| \omega \right\rangle
_{T_{1^{^{\prime }}}}-i\left| \omega +\delta \right\rangle _{T_1^{^{\prime
}}})+  \nonumber \\
&&\left| \omega \right\rangle _1\left| \omega \right\rangle _{4^{^{\prime
}}}(\left| \omega \right\rangle _{T_{1^{^{\prime }}}}+i\left| \omega +\delta
\right\rangle _{T_1})(\left| \omega \right\rangle _{T_2}+i\left| \omega
+\delta \right\rangle _{T_2^{^{\prime }}})  \nonumber \\
&=&\{[\left| \omega +\delta \right\rangle _{1^{^{\prime }}}\left| \omega
+\delta \right\rangle _4+\left| \omega \right\rangle _1\left| \omega
\right\rangle _{4^{^{\prime }}}]\left| \omega \right\rangle _{T_{1^{^{\prime
}}}}\left| \omega \right\rangle _{T_2}-  \nonumber \\
&&\lbrack \left| \omega +\delta \right\rangle _{1^{^{\prime }}}\left| \omega
+\delta \right\rangle _4+\left| \omega \right\rangle _1\left| \omega
\right\rangle _{4^{^{\prime }}}]\left| \omega +\delta \right\rangle
_{T_1}\left| \omega +\delta \right\rangle _{T_{2^{^{\prime }}}}- \\
&&i[\left| \omega +\delta \right\rangle _{1^{^{\prime }}}\left| \omega
+\delta \right\rangle _4-\left| \omega \right\rangle _1\left| \omega
\right\rangle _{4^{^{\prime }}}]\left| \omega +\delta \right\rangle
_{T_1}\left| \omega \right\rangle _{T_2}-  \nonumber \\
&&i[\left| \omega +\delta \right\rangle _{1^{^{\prime }}}\left| \omega
+\delta \right\rangle _4-\left| \omega \right\rangle _1\left| \omega
\right\rangle _{4^{^{\prime }}}]\left| \omega \right\rangle _{T_{1^{^{\prime
}}}}\left| \omega +\delta \right\rangle _{T_{2^{^{\prime }}}}]\}  \nonumber
\end{eqnarray}
From the equation above, if one photon enters AOM1 and at the same time
another photon enters AOM2 (refer to the Fig. 2 of the Ref. [1]), when one
of two detectors in modes T$_1$ or T$_{1^{^{\prime }}}$ and  one of the
other two detectors in modes T$_2$ or T$_{2^{\prime }}$ fire simultaneously,
photons 1 and 4 will be projected into a maximally frequency-entangled
state. The only difference between equation above and the Eq.(10) of the
Ref. [2] is a maximally entangled state $\left| \omega +\delta \right\rangle
_{1^{^{\prime }}}\left| \omega +\delta \right\rangle _4-\left| \omega
\right\rangle _1\left| \omega \right\rangle _{4^{^{\prime }}}$ can be
obtained with 50\% probability, which is not included in Eq.(10) of the
Ref.[2] because of improper transformation. So the main idea of our scheme
is correct, although it needs a small revision.

Similarly, AOM can also be used to create a Greenberger-Horne-Zeilinger
(GHZ) state. Comparing to Eq.(16) of the Ref.[2], the correct equation
should be (omit the constant factor)

\begin{eqnarray}
&&\lbrack \left| \omega \right\rangle _1\left| \omega +\delta \right\rangle
_{3^{^{\prime }}}\left| \omega \right\rangle _{4^{^{\prime }}}+\left| \omega
+\delta \right\rangle _{1^{^{\prime }}}\left| \omega \right\rangle
_{2^{^{\prime }}}\left| \omega \right\rangle _4]\left| \omega \right\rangle
_T+  \nonumber \\
&&i[\left| \omega \right\rangle _1\left| \omega +\delta \right\rangle
_{3^{^{\prime }}}\left| \omega \right\rangle _{4^{^{\prime }}}-\left| \omega
+\delta \right\rangle _{1^{^{\prime }}}\left| \omega \right\rangle
_{2^{^{\prime }}}\left| \omega \right\rangle _4]\left| \omega +\delta
\right\rangle _{T^{\prime }}.
\end{eqnarray}
Obviously, when one of detectors in modes T or T$^{^{\prime }}$ fires, Eq.
(4) will be projected into a GHZ state. The only difference is that two GHZ
states $\left| \omega \right\rangle _1\left| \omega +\delta \right\rangle
_{3^{^{\prime }}}\left| \omega \right\rangle _{4^{^{\prime }}}\pm \left|
\omega +\delta \right\rangle _{1^{^{\prime }}}\left| \omega \right\rangle
_{2^{^{\prime }}}\left| \omega +\delta \right\rangle _4$ can be created with
50\% probability respectively.

In conclusion. in spite of a small mistake we make, the main result of our
paper is correct, that is, AOM can be used to manipulate the
frequency-entangled state.

\end{document}